\documentstyle[aaspp4,tighten,psfig]{article}
\def\ptlandscape{\textwidth=9in \textheight=6.5in \hoffset=0in \voffset=-.25in}
\slugcomment{Accepted by Astronomical Journal: July 1998}
\begin{document}
\title{The HST Medium Deep Survey Cluster Sample: Methodology and Data}
 
\author{E. J. Ostrander, R. C. Nichol, K. U. Ratnatunga \& R. E. Griffiths}
\affil{Department of Physics, Carnegie Mellon University,
5000 Forbes Ave., Pittsburgh, PA--15213, USA}
 
\begin{abstract}
 
We present a new, objectively selected, sample of galaxy overdensities
detected in the Hubble Space Telescope Medium Deep Survey. These
clusters/groups were found using an automated procedure which involved
searching for statistically significant galaxy overdensities. 
The contrast of the clusters against the field galaxy population
is increased when morphological data is used to search around 
bulge--dominated galaxies.
In total, we present 92 overdensities above a probability threshold of
99.5\%.  We show, via extensive Monte Carlo simulations, that {\it at
least} $60\%$ of these overdensities are likely to be real clusters
and groups and not random line--of--sight superpositions of galaxies.
For each overdensity in the MDS cluster sample, we provide a richness
and the average of the bulge--to--total ratio of galaxies within each
system.  This MDS cluster sample potentially contains some of the most
distant clusters/groups ever detected, with about 25\% of the
overdensities having estimated redshifts $z {>\atop{\sim}} 0.9$.  We have made this
sample publicly available to facilitate spectroscopic confirmation of
these clusters and help more detailed studies of cluster and galaxy
evolution.
 
We also report the serendipitous discovery of a new cluster close on
the sky to the rich optical cluster CLl0016+16 at $z=0.546$. This new
overdensity, HST~001831+16208, may be coincident with both an X-ray
source and a radio source.  HST~001831+16208 is the third cluster/group
discovered near to CL0016+16 and appears to strengthen the claims of
Connolly et al.  (1996) of superclustering at high redshift.
\end{abstract}
 
 
\section{Introduction}
 
Clusters of galaxies$\footnote{Throughout this paper, we will use the
term ``cluster'' loosely to represent any statistical overdensity we
have detected.  This includes a range in richness from groups of
galaxies to rich Abell--like clusters.}$ are the largest known
gravitationally bound objects in the universe.  They are key tracers
of the large--scale structure in the universe (e.g. Postman, Huchra \&
Geller 1992) and provide an economical way of surveying large volumes
of space.  Furthermore, they represent the highest peaks in the
underlying mass density field and are therefore sensitive indicators
of the mean density of the universe {\it i.e.} $\Omega$ (see
Richstone, Loeb \& Turner 1992).  Clusters of galaxies have become one
of the favored tools of theoretical cosmologists as relatively simple
analytical methods can be developed to understand the distribution,
evolution and formation of clusters ({\it e.g.} \cite{PS74}).  In
addition to delineating the large--scale structure, clusters provide a
useful laboratory within which to study galaxy evolution as a function
of environment.
 
Classically, clusters/groups of galaxies have been found via visual
searches of optical galaxy catalogues for statistical excesses of
galaxies above the background population. These galaxy overdensities
are then characterized by richness and compactness criteria (a certain
surface density of galaxies is usually required). In addition, a
measure of the cluster's absolute brightness is obtained.  The first
major catalogues of clusters and groups of galaxies were produced by
Abell (1958) \& Zwicky et al. (1968) from visual scans of optical
photographic plates.  The Abell catalogue (Abell, Corwin \& Olowin
1989) is still widely used in present day astronomical research; it
contains more than 4000 nearby clusters, some as far out as
$z\simeq0.4$, and has been shown to be complete to $z\simeq0.15$
(Huchra et al. 1990).  The most widely used catalogue of groups of
galaxies is the Hickson catalogue of compact groups (median redshift
of $z=0.03$).  The reader is referred to an excellent recent review of
the Hickson catalogue -- and science extracted from it -- by Hickson
(1997).  Gunn, Hoessel \& Oke (1986) were the first to publish a
catalogue of distant optical clusters of galaxies, followed by Couch
et al. (1991).  These catalogues were also constructed from visual
scans of deep 4m photographic plates and combined contain 530
candidate clusters/groups out to $z\sim1$.
 
All these aforementioned optical cluster catalogues have two severe
limitations.  First, they are visually--based catalogues and
therefore, do not have well--determined selection functions. This
hinders a robust measurement of the volume sampled by these surveys.
This is a vital quantity which is needed for an unambiguous comparison
of the observations with theoretical predictions. Second, a caveat
which also applies to the present work, these optical clusters are
detected as overdensities in the projected distribution of galaxies
{\it i.e.}  in two dimensions. This can result in ``phantom'' clusters
which appear to be clusters on the sky, but are, in reality,
superpositions of galaxies along the line--of--sight. Numerical
simulations of the contamination by projection effects suggest that up
to 50\% of all Abell clusters could be spurious (Lucey 1983; Frenk et
al. 1990; van Haarlem, Frenk \& White 1997)
 
In recent years, the situation regarding optical catalogues of
clusters has changed.  The Palomar Distant Cluster Survey (PDCS,
Postman et al. 1996) is the first fully automated survey for optical
clusters of galaxies and takes advantage of large CCD cameras on
4-metre class telescopes.  This makes it possible to survey large
areas of the sky to faint magnitude limits.  The PDCS covers a total
area of $5 {\rm deg^2}$ in two bands -- $V_4$ and $I_4$ -- to a
completeness magnitude of $I\simeq22$.  Postman et al. used a
matched--filter cluster--finding algorithm that detected 107 clusters
in total; 79 of which formed a complete sample of clusters in the
estimated redshift range of $0.2<z<0.6$. Via extensive Monte Carlo
simulations, the PDCS has a well--known selection function and
therefore, the volume sampled by the survey has been determined.
These simulations were also used to determine the completeness of the
survey and the likely contamination rate by ``phantom clusters''
(estimated to be $<30\%$).  In addition to the PDCS, there are several
other examples of the use of objective criteria to find and classify
optical clusters (see Lumsden et al. 1992; Lidman \& Peterson 1996;
Kawasaki et al. 1997; Kepner et al. 1998; Kodama, Bell \& Bower 1998).
 
In this paper, we present a new catalogue of distant optical clusters
selected objectively from the Hubble Space Telescope (HST) Medium Deep
Survey (MDS).  The motivation for this new survey is to increase the
total number of potentially high redshift clusters.  This survey will
be known as the MDS Cluster Sample. In Section 2, we briefly summarize
the overall MDS methodology and catalog.  In Section 3, we discuss the
cluster--finding algorithm while in Section 4 we outline the sample of
92 overdensities in the MDS cluster sample.  In Section 5, we present
the simulations we have performed to assess the robustness of the
clusters selected, while in Section 6 we discuss the selection
function of the MDS cluster sample.  Section 7 we give our discussions
and conclusions.
 
\section{The HST Medium Deep Survey}
 
The MDS is a long--term project dedicated to extracting as much
scientific information as possible out of the large amount of parallel
data taken by the HST Wide Field and Planetary Camera (WFPC2). For a
complete summary of the methodology used in the MDS, the reader is
referred to Griffiths et al. (1994) but for completeness, we review the
salient points here.
 
To date, the MDS contains over 800 WPFC2 fields scattered over the
whole sky, of which over 500 of the better quality fields have been
fully analyzed.  The analysis involves a semi-automated data reduction
pipeline that is designed to take the raw HST WPFC2 data and produce
calibrated catalogues of model--fitted galaxies. The first step
involves dark and bias subtraction followed by flat--fielding and hot
pixel removal.  The images are then added to help remove cosmic--ray
events. The reader is referred to Ratnatunga et al. (1994) for a full
review of all these procedures.
 
The second step is object detection which involves a local
peak--finding algorithm that locates contiguous pixels one--sigma
above a local sky determination.  These detected objects are then
visually inspected. The third and final step in the MDS analysis is a
maximum likelihood model fit for the morphology of the objects in each
field. This involves a 2--dimensional Disk+Bulge decomposition of
these faint undersampled images and has been shown to provide an
unbiased estimate of the real morphology of the sources down to F814W
= 22.7, F606W = 23.3 and F450W =23.1 in a one hour exposure.  In this
way, the MDS provides reliable morphologies for each detected object
{\it e.g.} stellar, faint galaxy, bulge--like galaxy, disk--like
galaxy or a bulge+disk galaxy.  The reader is referred to Ratnatunga,
Griffiths \& Ostrander (1998) for a detailed explanation of this
fitting procedure (or see http://astro.phys.cmu.edu/mds/mle/).
 
For the work discussed in this paper, we do not use all available
fields in the MDS. A summary of the fields we have used and the
composition of their galaxy morphologies is given in Table 1. These
MDS fields are scattered over the whole sky but we have excluded all
fields at low galactic latitude, $|b|\le16^{\circ}$, to help minimize
false detections due to the increased surface density of stars.  We
have also excluded some high galactic latitude fields dominated by
stars ({\it i.e.} LMC), nearby galaxies (to avoid globular clusters)
and MDS fields where the primary target was a known cluster.
 
%


\tablenum{1}


\begin{table}[tb]
\begin{center}
\caption{Number of HST WFP2 fields used in the 
MDS cluster sample. Also presented is the effective
area (${\rm arcmins^2}$) -- accounting for field overlaps and
edge effects (see text) --
as a function of HST passband. 
The number of galaxies, regardless of their morphology, 
in each passband is given along
with the number of disk--dominated and bulge--dominated
galaxies (see text for definitions). The total number
of galaxies far exceeds the number of bulge and disk--dominated
galaxies because it includes lower signal--to--noise 
galaxies that were not classified.
\label{mdsfields}
}

\begin{tabular}{lrrrrrr}\hline\hline
Filter & No. & Unique &Total & \multicolumn{2}{c}{Morphology} \\
       & Fields & Area & Galaxies   & Bulge & Disk \\ \hline
F450W (b) & 29 & 119.6 & 10070 & 619 & 3595 \\
F606W (v) & 251 & 1062.0 & 69313 & 5917 & 27670 \\
F814W (i) & 319 & 1285.1 &89790 & 7118 & 36649 \\ \hline \hline
\end{tabular}
\end{center}
\end{table}


Throughout this paper, we use the term ``bulge dominated'' to describe
a galaxy that has a $\ge50\%$ bulge model component in the MDS 2D
morphological model fit as discussed above and in detail in Ratnatunga
et al. (1994, 1998). In other words, the dominant component of the
galaxy profile is a bulge as seen in ellipticals and S0
galaxies. Likewise, we use the term ``disk dominated'' to describe a
galaxy that has $>50\%$ disk component in the 2D MDS morphological
model fit.

\begin{figure}[tp]
\centerline{\psfig{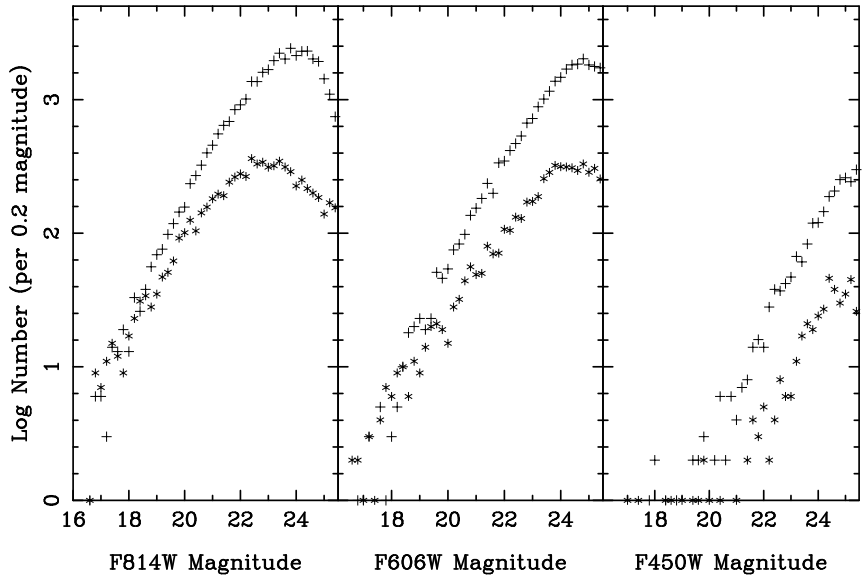}}
\caption{
The number of galaxies used in this paper as a function of passband,
magnitude and morphology.  This figure is not meant to be used as the
number--magnitude relationship for the whole MDS survey as that has
been published elsewhere.  The $+$ symbols represent galaxies
classified as disk dominated and $\star$ represent galaxies classified
as bulge dominated. See text for the definition of these two
morphological classifications.
\label{nmplot}
}
\end{figure}
 
In Fig.~\ref{nmplot}, we present the number of galaxies used in this
paper as a function of magnitude, passband and galaxy morphology.
This is not meant to be used as a number--magnitude relationship for
the whole MDS galaxy survey since this has been published elsewhere
(Roche et al. 1998).  In Table 2, we present the differential areal
coverage, as a function of the completeness limit of each field, for
the MDS cluster sample.  The field completeness limit is calculated
based on the exposure time, number of exposures and the local sky
background (see Ratnatunga et al. 1998).  The total amount of area
covered by the MDS cluster survey, after the removal of overlaps
between neighbouring fields, is given in Table 1.  These data are
vital to the usefulness of the MDS cluster sample as they will
be needed for any estimate of the volume sampled by our
cluster survey.



\tablenum{2}

\begin{table}[tb]
\begin{center}
\caption{The differential area of the MDS cluster sample 
as a function of 
limiting magnitude (in ${\rm arcmin^{2}}$). We have accounted
for edge effects (see text)
\label{tablearea}
}
\begin{tabular}{cr|cr|cr} \hline \hline
\multicolumn{2}{c|}{b-band}&
\multicolumn{2}{c|}{v-band}&
\multicolumn{2}{c}{i-band}\\
mag & area &
mag & area &
mag & area \\ \hline
   24.625&   8.54 &    23.625&       12.81&       22.625&       12.81\\    
   24.875&   12.81&    23.875&       15.54&       22.875&       11.25\\    
   25.125&   51.24&    24.125&       95.97 &    23.125&       66.52\\   
   25.375&   29.89&    24.375&       160.17 &    23.375&       106.71\\     
   25.625&   12.81&   24.625&       193.98&    23.625&       147.74 \\    
   26.625&   4.27 &   24.875&       106.68&    23.875&       178.65\\    
         &         &   25.125&       264.30&   24.125&       212.72\\        
         &         &   25.375&       76.86&     24.375&       330.82\\      
         &         &   25.625&       55.51&      24.625&       124.32\\     
         &         &   25.875&       8,54&       24.875&       68.32\\    
         &         &   27.125&       4.27&       25.125&       4.27\\    
         &         &         &            &       26.125&       4.27\\ \hline    
\end{tabular}
\end{center}
\end{table}

Table 2 was computed by summing the number of fields as a function of
their limiting magnitude and accounting for overlap between adjacent
fields.  The degree of overlap was small; for the F814W data only
$2\%$ of the area overlapped with an adjacent MDS field.  We used the
effective total area as defined in Section 6 for the available search
area per WFPC2 field ($4.27\,\,{\rm arcmins^{2}}$).
 
\section{Construction of the Cluster Sample}
 
\subsection{Overview}
 
The requirements of a cluster--finding algorithm are clear; it must
maximize the number of true clusters detected, in a noisy correlated
background, while minimizing the number of false detections.
Moreover, it should have as simple a definition as possible, allowing
the selection function of the cluster survey to be well understood.
Ideally, the algorithm would make as few assumptions as possible about
the clusters it is trying to detect to minimize possible biases.
 
The PDCS used a matched--filter technique which involves filtering the
data with a physical model of a cluster. This helps to increase the
contrast of any cluster as a function of redshift and radius.
However, it does impose a particular cluster model on the data.  Other
techniques have included simply peak--finding within galaxy catalogues
(Lumsden et al. 1992), a maximum likelihood method similar to the PDCS
algorithm (Kawasaki et al. 1997) and a wavelet--based method (Slezak,
Bijaoui \& Mars 1990).
 
The MDS is a unique database of galaxies upon which to build a new
search for distant clusters of galaxies.  This is primarily because it
is the largest collection of galaxy morphologies in existence and we
can utilize this morphological information to develop a simple and
effective cluster--finding algorithm taking advantage of the
density--morphology relationship (Dressler 1980).  It has been shown
that clusters and groups are dominated by elliptical galaxies whereas
the field galaxy population is predominantly made up of spirals. It is
therefore possible to greatly enhance the contrast of distant clusters
of galaxies against the field population with morphological data.
 
In addition to the morphological information, the MDS includes some of
the deepest photometric data ever obtained, meaning that the MDS
cluster sample has, on average, a fainter limiting magnitude than any
previous search for distant clusters (see Fig.~\ref{nmplot}), while
still covering a reasonable area of the sky ($0.4 {\rm deg^{2}}$ in
F814W, see Table 2).  We have included in the search the Hubble Deep
Field (HDF; Williams et al. 1996) and the Groth--Westphal strip, since
these fields were not selected for any particular target and
therefore, should be representative of the galaxy population.
Moreover, it allows us to statistically quantify the existence of any
galaxy overdensities in these very deep images.
 
We note that any cluster/group search based on the morphologies of the
galaxies is potentially biased due to evolutionary effects.  This may
limit the usefulness of the catalogue; for example, detailed studies
of the morphological constitution of these MDS distant clusters could
be undermined (see Section 6).  We believe however, that our algorithm
may be robust against such concerns for two reasons.  First, we only
require one bulge dominated galaxy per cluster {\it i.e.} one
elliptical or S0.  Second, bright cluster ellipticals are thought to
have formed at very high redshift ($z>1$) and have evolved passively
since then (see Dickinson 1997).  Finally, the number of irregular
galaxies increases at high redshift which potentially biases our
result as we do not search around such galaxies. However, the total
number of irregulars detected in the Hubble Deep Field is only $\sim
25\%$ and therefore, we do not envisage this to be a large effect for
our shorter exposure MDS fields.
 
\subsection{The Mechanics of the Algorithm}
 
Here we discuss the details of the cluster--finding algorithm.  The
original plan for our cluster--finding algorithm was to base our
search for overdensities on the photometric redshifts of galaxies,
allowing us to scale a fixed metric aperture, and a fixed absolute
magnitude width, with this redshift.  However, our initial
investigations of such an algorithm showed that the errors on the
photometric redshifts, because of our limited color data, were too
high.  Moreover, in the redshift range we are probing with this MDS
data, $z\simeq0.5$ to $z\simeq1$, the angular diameter--redshift
relationship is almost flat and therefore, there is little need to
scale a metric aperture with redshift.
 
Our algorithm was based on the search for statistically significant
galaxy overdensities in an array of different apertures centered on
each bulge dominated galaxy in the MDS.  We vary both the radius and
the magnitude range of these search apertures (see below) to increase
our probability of finding clusters over a large range in redshift.
This approach also means that our selection is less biased towards
finding clusters with a particular radial profile and/or galaxy
luminosity function.
 
The angular size of the apertures were varied from 10 to 35 arcseconds
in radius, in steps of 5 arcseconds, which
corresponds to a range of $80\,h^{-1}{\rm \,kpc}$ to $285\,h^{-1}{\rm
\,kpc}$ in metric diameters for a median redshift of $\simeq0.75$ (we
use $q_o=0.5$ and ${\rm H_o}=100\,{\rm km\,s^{-1}\,Mpc^{-1}}$, or
$h=1$ throughout).  The magnitude range of the apertures is fixed at 2
magnitudes ($0.5\le m_c < 1.5$) but the central magnitude ($m_c$) is
varied, in steps of 0.25 magnitude, from 16.5 magnitudes to 1.5
magnitudes brighter than the completeness limit of each individual
field.  The bright magnitude limit was chosen based on
Fig.~\ref{nmplot} which shows there are very few galaxies brighter
than this, in all 3 HST passbands, within the MDS database.
Therefore, the number of apertures used per `seed' galaxy (the galaxy
at the center of the aperture) varies depending on the field, but
typically $\sim60$ different radius/magnitude apertures are applied.
We excluded from the analysis apertures that had less than 75\% of
their area within the WFPC2 image.
 
The algorithm used for the MDS cluster sample was broken into two
stages.  The first stage involved processing all galaxies regardless
of their morphology and passband as given in Table 1.  For each
galaxy, we counted the number of galaxy neighbors in the array of
apertures discussed above which resulted in a set of distribution
functions that show the frequency of neighboring galaxies as both a
function of search radius and magnitude.  These functions form the
basis for our overdensity search and act as our null hypothesis.  They
are our determination of the background galaxy population against
which we can test the statistical significance of any observed
overdensity.  Simply put, these distributions represent our
measurement of the galaxy angular correlation function.  The second
stage of the construction of the MDS cluster sample is very similar to
the first stage, however, in this stage we only placed apertures on
bulge dominated galaxies and once again counted the number of
neighboring galaxies.
 
We are now faced with a problem that is common to all objective
searches for clusters.  We need to determine the threshold above which
a galaxy overdensity enters our sample.  Such a threshold can either
be determined empirically by maximizing the number, or type, of
cluster one wants to find, or one can simply set an arbitrary
threshold.  Irrespective of the actual threshold, it is very important
to document the decision to provide future users of the catalogue a
clear view of what was included and discarded. This will greatly aid
comparison with theoretical models.
 
For the MDS cluster sample, we empirically selected our thresholds --
which are richness cutoffs -- from the distributions constructed in
the first stage discussed above.  For each combination of aperture
size and magnitude range, we computed the galaxy richness that
corresponds to the 99.5 percentile in each distribution {\it i.e.} the
galaxy richness of an overdensity above which it is in the top 0.5\%
of all galaxy overdensities observed in the MDS.  An example of these
galaxy richness cutoffs, as a function of radius and magnitude, is
given in Table 3.  In summary, we have presented all groups/clusters
that are in the tail of the distribution.



\tablenum{3}

\begin{table}[thp]
\begin{center}
\caption{The galaxy richness cutoffs determined
in stage 1 of the construction of the MDS cluster sample
and corresponding to the 99.5\% probability
threshold.
These cutoffs are given
as a function of the aperture radius and the 
magnitude of the `seed' galaxy (or $m_c$) in the F814W filter.
\label{cutoffs_i}
}
\begin{tabular}{c|rrrrrr}\hline \hline
    &\multicolumn{6}{c}{radii (arcseconds)}\\ \hline 
$m_c$ & 10 & 15 & 20 & 25 & 30 & 35 \\ \hline
  17.00&   2.37&   2.63&   3.27&   3.40&   3.66&   3.86\\
  17.25&   2.56&   2.63&   3.27&   3.40&   4.00&   4.24\\
  17.50&   2.55&   4.41&   5.45&   5.94&   6.38&   5.86\\
  17.75&   2.55&   4.41&   4.65&   4.85&   5.19&   5.59\\
  18.00&   2.35&   3.38&   3.71&   4.27&   5.09&   6.12\\
  18.25&   2.55&   3.43&   4.08&   4.55&   5.54&   7.01\\
  18.50&   2.99&   3.55&   4.08&   4.99&   5.99&   7.16\\
  18.75&   3.09&   3.59&   4.99&   5.38&   6.39&   7.47\\
  19.00&   3.99&   6.00&   5.99&   6.99&   9.05&   9.25\\
  19.25&   3.74&   5.99&   5.99&   6.99&   8.99&   9.44\\
  19.50&   4.99&   8.99&   8.99&  10.99&  12.99&  14.17\\
  19.75&   4.99&  10.99&  11.99&  13.13&  16.99&  18.08\\
  20.00&   4.99&   7.19&   9.17&  12.99&  17.26&  20.03\\
  20.25&   5.99&   8.54&   9.73&  11.46&  14.79&  17.39\\
  20.50&   4.99&   7.99&   9.99&  11.99&  15.01&  18.02\\
  20.75&   5.99&   8.65&  11.99&  13.99&  17.99&  20.72\\
  21.00&   5.99&   8.65&  10.99&  14.02&  17.00&  20.54\\
  21.25&   5.99&   8.99&  11.99&  15.37&  19.13&  22.93\\
  21.50&   6.43&   9.93&  12.70&  15.75&  20.38&  25.73\\
  21.75&   6.78&  10.01&  13.99&  17.99&  22.94&  29.33\\
  22.00&   7.04&  11.11&  15.20&  20.40&  26.46&  32.10\\
  22.25&   7.99&  12.07&  17.41&  23.86&  31.47&  40.79\\
  22.50&   9.13&  14.00&  19.99&  27.37&  36.28&  44.46\\
  22.75&   9.85&  14.99&  21.11&  30.30&  40.32&  51.54\\
  23.00&  10.19&  16.95&  24.81&  36.13&  47.06&  63.29\\
  23.25&  10.41&  19.40&  30.14&  41.99&  57.08&  76.90\\
  23.50&  12.99&  22.99&  34.07&  47.11&  65.29&  89.33\\
  23.75&  13.22&  23.80&  35.29&  50.18&  70.64&  92.83\\
  24.00&  14.73&  24.95&  37.95&  52.06&  69.20&  94.42\\
  24.25&  17.00&  29.15&  45.23&  60.53&  80.87& 107.01\\
  24.50&  17.55&  32.45&  51.03&  68.28&  88.01& 112.04\\
  24.75&  18.13&  32.26&  50.19&  72.22&  99.81& 129.19\\ \hline
  \hline
\end{tabular}
\end{center}
\end{table}

This percentile cutoff (99.5\%) was determined from a visual
inspection of the MDS cluster detections.  This method does not
undermine the objectivity of the cluster selection since, above this
chosen percentile threshold, all detected overdensities, regardless of
their subjective visual appearance, are included.  The threshold
corresponds to an arbitrary richness (or mass) cutoff that can be
solidified later via multi--object spectroscopy.
 
At the end of the second stage, nearly a thousand cluster candidates
were detected above the richness cutoffs discussed above and shown in
Table 3.  They were separated into 3 catalogues, one for each of the
HST passbands.  This large number of candidate clusters is simply a
reflection of the same real clusters being found many times over as
each bulge dominated galaxy in a cluster can provide a potential
detection.  Overlaps were removed by sorting the catalogues, as a
function of passband, and finding any candidates that had the same
central right ascension and declination within a radius of 30
arcseconds. For these common candidates, the one with the highest
detection percentile was chosen as the main cluster detection.  If the
candidates had the same detection percentile, then the one with the
highest galaxy richness was taken.  Once the cluster catalogues for
the individual passbands were sorted, they were combined using the
same sorting algorithm {\it i.e.} if two candidates, in different
passbands, had the same central right ascension and declination within
a radius of 30 arcseconds, the candidate with the highest detection
percentile was chosen. If the candidates had the same detection
percentile, then the one with the highest galaxy richness was taken.
 
This sorting substantially reduced the size of cluster detections
leaving only 126 candidate groups/clusters.  All of these systems were
visually inspected which showed that several of these candidates were
still the same real cluster but had been found at different $m_c$
magnitudes.  These remaining duplicate cluster candidates were removed
by hand leaving the candidate with the highest percentile or the
largest galaxy richness (if the duplicates had the same percentile).
We were careful not to remove candidates that were close on the sky
but appeared to be two separate entities {\it i.e.} a distant cluster
behind a nearby group (see Section 6).  This procedure resulted in a
catalogue of 92 unique groups/clusters.
 
\section{Results}
 
Table 4 contains the 92 overdensities detected above the 99.5\%
probability threshold discussed in Section 3.2.  In this table we
present a unique name (column 1) and the right ascension and
declination (J2000) of the `seed' galaxy for the cluster in columns 2
\& 3.  The galaxy richness within the detection aperture is given in
column 4. This richness has been corrected for any area of the
detection aperture outside the field {\it i.e.} it is scaled assuming
a constant surface density of galaxies to give the expected galaxy
richness if the aperture was fully within the MDS field. This explains
the non--integer richnesses in Table 4.  The galaxy richnesses are
however not corrected for background or foreground galaxy
contamination. There were three reasons for this decision.  First,
neither a global or a local contamination correction would have worked
satisfactorily. This is because the field--to--field variations in the
MDS are large thus negating a global approach. Meanwhile a local
correction would have been hindered by the fact that in most cases,
the presence of a cluster in a given field would have significantly
skewed the galaxy counts in that field, thus leading to an over
estimate for the correction.  Second, $\sim40\%$ of our overdensities
have low measured richnesses, {\it i.e.} less than 10 members (because
of the small apertures), which raises the issue of small number
statistics. Finally, the MDS survey is unique because of the
morphological information it contains and it becomes increasingly
difficult to make a statistical correction for contamination as a
function of galaxy morphology.  As a first order correction to the
richness, the reader may wish to multiply the richness by the
bulge--to--total ratio of galaxies (Column 9 of Table 4, see below)
thus obtaining the number of bulge dominated galaxies in each
cluster/group. This may be a more physical richness estimate because
of the density--morphology relationship {\it i.e.} elliptical galaxies
populate the cores of clusters.



{\footnotesize

\tablenum{4}

\begin{table}[tp]
\def\-{$-$}

\caption{The 92 
MDS clusters detected above a 99.5\% probability 
threshold as discussed in Section 3.2.
\label{datatable}
}
{\footnotesize
\begin{tabular}{lrrrrrrrrrrr|l}\hline\hline
Name & RA & Dec & ${\rm N_{gal}}$ & $\%$ & ${\rm N_{gal\%}}$ & rad & mag & B/T & F & clim & G & Primary Target   \\ \hline
HST 001548\-16200 &  3.95021 &\-16.33396 &23.00 &99.53& 22.94& 30& 21.75& 0.29&i& 23.92 &3& L722-22-0002     \\
HST 001557\-16184 &  3.98791 &\-16.30701 &10.52 &99.93&  7.09& 10& 22.75& 0.33&v& 24.23 &1& L722-22-0014     \\
HST 001831+16207 &  4.62980 & 16.34527 &10.00 &99.93&  7.99& 10& 22.25& 0.57&i& 24.03 &2& QSO0015+162      \\
HST 002013+28368 &  5.05543 & 28.61357 &10.00 &99.99&  6.79& 10& 21.75& 0.38&i& 24.59 &1& QSO0020+287      \\
HST 002458\-27167 &  6.24412 &\-27.27936 & 9.00 &99.99&  7.05& 10& 22.00& 0.20&i& 24.75 &2& LHS1070-B        \\
HST 004838+85109 & 12.16184 & 85.18242 & 9.45 &99.50&  9.45& 35& 19.25& 0.35&i& 22.67 &4& NGC188-AA        \\
HST 004933\-52046 & 12.38972 &\-52.07701 &10.56 &99.55& 10.19& 10& 23.00& 0.25&i& 24.45 &3& BPM16274         \\
HST 005017\-52122 & 12.57334 &\-52.20405 &18.20 &99.61& 17.99& 25& 21.75& 0.15&i& 23.84 &4& BPM16274         \\
HST 005020\-52113 & 12.58685 &\-52.18839 & 5.02 &99.99&  3.97& 10& 20.75& 0.50&v& 24.28 &2& BPM16274         \\
HST 005807\-28106 & 14.53048 &\-28.17803 & 6.00 &99.52&  6.00& 15& 22.25& 0.17&b& 24.91 &2& SGP1-10          \\
HST 010957\-02276 & 17.48903 & \-2.46027 & 5.34 &99.99&  5.25& 35& 21.00& 0.21&b& 24.98 &4& Q0107-025B       \\
HST 011704\-08386 & 19.26687 & \-8.64425 & 7.78 &99.99&  7.28& 30& 19.75& 0.44&v& 24.06 &2& Q0114-089        \\
HST 012006+21273 & 20.02547 & 21.45598 &14.96 &99.69& 14.00& 15& 22.50& 0.12&i& 24.05 &4& 0117+213         \\
HST 013835+33043 & 24.64890 & 33.07246 & 8.25 &99.65&  7.99& 10& 23.00& 0.50&v& 24.50 &3& 0134+329INCA221  \\
HST 020959\-39354 & 32.49727 &\-39.59015 &12.00 &99.82& 11.00& 25& 21.50& 0.61&v& 23.00 &3& Q0207-398        \\
HST 021002\-39356 & 32.50857 &\-39.59459 &16.65 &99.76& 15.00& 15& 23.50& 0.30&v& 25.05 &4& Q0207-398        \\
HST 021005\-39350 & 32.52127 &\-39.58443 &16.88 &99.76& 15.74& 35& 21.50& 0.57&v& 25.05 &4& Q0207-398        \\
HST 022548+27547 & 36.45216 & 27.91263 & 6.51 &99.79&  6.00& 20& 19.25& 0.49&i& 23.84 &2& RWTRI-GSC-2      \\
HST 035528+09435 & 58.86759 &  9.72577 &13.00 &99.99&  7.99& 10& 22.25& 0.21&i& 23.41 &1& HZ4              \\
HST 035531+09441 & 58.88189 &  9.73624 & 8.26 &99.93&  6.43& 10& 21.50& 0.31&i& 23.90 &4& HZ4              \\
HST 035535+09433 & 58.89678 &  9.72319 & 6.05 &99.90&  5.10& 15& 20.75& 0.42&v& 25.01 &1& HZ4              \\
HST 045648+03529 & 74.20113 &  3.88260 & 7.15 &99.56&  7.05& 10& 22.00& 0.50&i& 24.39 &3& PKS0454+039      \\
HST 051909\-45493 & 79.78752 &\-45.82233 & 6.61 &99.72&  6.05& 15& 21.00& 0.50&v& 25.34 &4& PKS0518-45       \\
HST 051910\-45510 & 79.79485 &\-45.85162 &21.20 &99.70& 19.29& 30& 22.25& 0.20&v& 25.34 &4& PKS0518-45       \\
HST 072049+71089 &110.20662 & 71.14918 & 9.98 &99.60&  9.45& 35& 19.25& 0.49&i& 23.66 &4& 0716+714INCA221  \\
HST 072442+60316 &111.17674 & 60.52826 &15.00 &99.70& 14.00& 15& 22.50& 0.25&i& 24.07 &2& STAR-72553+60    \\
HST 072455+60313 &111.23310 & 60.52233 &11.14 &99.53& 11.07& 15& 22.75& 0.14&v& 24.57 &2& STAR-72553+60    \\
HST 074239+49428 &115.66546 & 49.71365 &13.98 &99.93& 11.11& 15& 22.00& 0.30&i& 24.25 &2& MRK79            \\
HST 075047+14412 &117.69848 & 14.68775 &23.41 &99.65& 22.04& 20& 23.50& 0.14&v& 24.88 &2& STAR-75117+14    \\
HST 095007+39248 &147.53330 & 39.41372 &11.46 &99.50& 11.45& 35& 20.75& 0.34&v& 25.08 &3& PG0947+396       \\
HST 095012+39244 &147.55212 & 39.40683 &26.76 &99.74& 25.74& 35& 21.50& 0.23&i& 24.18 &2& PG0947+396       \\
HST 100456+05151 &151.23347 &  5.25174 & 5.05 &99.60&  5.04& 15& 22.00& 0.43&b& 25.03 &3& PG1001+054       \\
HST 102722+03257 &156.84344 &  3.42847 &14.32 &99.72& 13.99& 20& 21.75& 0.23&i& 23.62 &1& CH02             \\
HST 111744+44177 &169.43644 & 44.29603 & 5.52 &99.99&  5.52& 20& 21.50& 0.20&b& 24.98 &3& PG1114+445       \\
HST 112125\-24558 &170.35809 &\-24.93087 &21.33 &99.60& 20.39& 30& 21.50& 0.38&i& 23.93 &4& HD98800-5-REF    \\
HST 115027+28496 &177.61418 & 28.82765 &10.89 &99.77& 10.00& 10& 23.75& 0.29&v& 25.15 &2& RE1149545+284512 \\
HST 121111+39273 &182.79823 & 39.45605 &29.04 &99.95& 26.99& 20& 24.00& 0.26&v& 25.49 &3& NGC4151-PO       \\
HST 121754+50123 &184.47775 & 50.20501 &17.06 &99.64& 16.96& 15& 23.00& 0.45&i& 24.64 &4& HS-1216+5032B    \\
HST 122332+15518 &185.88700 & 15.86494 &23.53 &99.76& 21.12& 20& 22.75& 0.07&i& 24.45 &2& SN1979C          \\
HST 122355+15495 &185.98101 & 15.82555 & 8.00 &99.80&  7.05& 10& 22.00& 0.30&i& 24.27 &2& 1220+160         \\
HST 123155+14163 &187.98116 & 14.27305 &46.65 &99.55& 46.30& 30& 23.75& 0.13&v& 25.24 &2& AL-COM           \\
HST 123639\-00417 &189.16295 & \-0.69544 & 9.78 &99.91&  7.99& 10& 22.25& 0.17&i& 24.17 &2& QNY1-32          \\
HST 123640+62111 &189.16921 & 62.18616 &24.95 &99.67& 23.86& 25& 22.25& 0.29&i& 24.08 &3& HDF              \\
HST 123649+62132 &189.20545 & 62.22007 & 7.82 &99.99&  7.82& 30& 22.00& 0.41&b& 26.20 &2& HDF              \\
HST 125015+31254 &192.56439 & 31.42368 &34.00 &99.99& 32.20& 30& 23.75& 0.16&b& 25.53 &1& CSO173           \\
HST 125651+22062 &194.21284 & 22.10452 & 9.13 &99.50&  9.13& 10& 23.50& 0.25&v& 25.32 &2& GD153            \\
HST 125655+22057 &194.23291 & 22.09520 & 7.05 &99.50&  7.05& 10& 22.00& 0.67&i& 24.60 &2& GD153            \\
HST 133605+51494 &204.02123 & 51.82415 &29.18 &99.99& 25.92& 30& 22.75& 0.32&v& 25.12 &3& UX-UMA           \\
HST 133617\-00526 &204.07282 & \-0.87794 &13.59 &99.99&  9.72& 30& 20.75& 0.48&v& 24.29 &1& QSO-133647-004858\\
HST 140428+43196 &211.11817 & 43.32792 & 5.59 &99.99&  5.59& 35& 17.75& 0.37&i& 24.75 &1& IR1402+43        \\
HST 141506+52015 &213.77739 & 52.02592 & 7.26 &99.61&  7.05& 10& 22.00& 0.33&i& 24.30 &3& 141816+523430    \\
HST 141610+52123 &214.04513 & 52.20645 & 5.89 &99.70&  5.23& 20& 20.50& 0.23&v& 25.04 &2& 141613+521222    \\
HST 141612+52133 &214.05358 & 52.22236 & 5.22 &99.83&  5.07& 20& 20.25& 0.41&v& 25.05 &4& 141619+521332    \\
HST 141613+11316 &214.05788 & 11.52808 &15.59 &99.85& 14.72& 30& 21.75& 0.21&v& 25.19 &3& Q1413+117-D      \\
HST 141618+52138 &214.07730 & 52.23165 &18.85 &99.99& 15.45& 25& 22.25& 0.45&v& 25.05 &1& 141619+521332    \\
HST 141624+52155 &214.10012 & 52.25913 &11.00 &99.80&  9.86& 10& 22.75& 0.17&i& 25.06 &3& 141626+521442    \\
HST 141637+52163 &214.15461 & 52.27239 &10.11 &99.61&  9.65& 20& 21.75& 0.69&v& 25.06 &3& 141632+521552    \\
HST 141638+52165 &214.16227 & 52.27623 &18.53 &99.67& 17.99& 25& 21.75& 0.10&i& 24.37 &3& 141638+521702    \\ \hline\hline
\end{tabular}
}
\end{table}

}

{\footnotesize
\tablenum{4}


\begin{table}[th]
\def\-{$-$}
\ptlandscape
\footnotesize{
\caption{Continued}
\begin{tabular}{lrrrrrrrrrrr|l}\hline\hline
Name & RA & Dec & ${\rm N_{gal}}$ & $\%$ & ${\rm N_{gal\%}}$ & rad & mag & B/T & F & clim & G & Target  \\ \hline
HST 141653+52210 &214.22210 & 52.35111 &13.50 &99.83& 11.99& 25& 20.50& 0.34&i& 24.37& 1& 141658+522032 \\
HST 141654+52189 &214.22910 & 52.31604 &11.00 &99.90& 10.00& 20& 20.50& 0.32&i& 25.06& 2& 141651+521922 \\
HST 141727+52267 &214.36530 & 52.44579 & 6.85 &99.53&  6.79& 10& 21.75& 0.24&i& 24.37& 2& 141731+522622 \\
HST 143518+24589 &218.82745 & 24.98225 &13.26 &99.99& 10.00& 10& 23.75& 0.11&v& 24.41& 1& G166-37       \\
HST 144152\-17175 &220.46995 &\-17.29237 &19.09 &99.81& 17.41& 20& 22.25& 0.28&i& 24.52& 3& NGC5728-EELR  \\ 
HST 150620+01448 &226.58519 &  1.74697 & 5.11 &99.50&  5.10& 15& 20.75& 0.26&v& 25.21& 4& NGC5845-FOS   \\
HST 150621+01431 &226.59099 &  1.71872 & 7.00 &99.50&  7.00& 25& 19.25& 0.77&i& 24.05& 3& NGC5845-FOS   \\
HST 151940+23524 &229.91835 & 23.87468 &22.78 &99.93& 20.41& 25& 22.00& 0.33&i& 24.57& 2& LB9605-NEW    \\
HST 162413+48077 &246.05638 & 48.12981 &12.00 &99.90&  9.86& 10& 22.75& 0.33&i& 24.25& 2& GL623-5-REF   \\
HST 162413+48078 &246.05653 & 48.13023 & 6.35 &99.54&  6.00& 10& 21.25& 0.71&i& 23.42& 3& AC+48D1595-89 \\
HST 163141+37375 &247.92343 & 37.62648 &14.26 &99.70& 13.99& 20& 21.75& 0.45&i& 24.20& 2& PG1630+377    \\
HST 171220+33354 &258.08595 & 33.59067 &11.00 &99.80& 10.41& 10& 23.25& 0.11&i& 24.66& 3& V795-HER      \\
HST 171223+33371 &258.09734 & 33.61898 & 6.00 &99.70&  5.80& 10& 22.00& 0.72&v& 25.57& 4& V795-HER      \\
HST 173638+68251 &264.16141 & 68.41998 &13.21 &99.88& 11.51& 25& 21.75& 0.19&v& 24.27& 1& BD+68D946     \\
HST 175525+18182 &268.85509 & 18.30430 &21.85 &99.99& 17.99& 25& 21.75& 0.25&i& 24.68& 1& NGC6500       \\
HST 180746+45599 &271.94548 & 45.99925 & 6.00 &99.60&  5.99& 10& 21.00& 0.82&i& 24.55& 3& DQ-HER        \\
HST 193810\-46205 &294.54349 &\-46.34241 &10.10 &99.62&  9.86& 10& 22.75& 0.29&i& 24.46& 3& QS-TEL        \\
HST 193928\-46139 &294.86894 &\-46.23330 & 7.91 &99.89&  6.43& 10& 21.50& 0.29&i& 24.41& 2& STAR-193835-46\\
HST 194752\-41520 &296.96836 &\-41.86737 & 7.00 &99.80&  6.00& 10& 21.25& 0.31&i& 23.86& 2& V3885-SGR     \\
HST 194754\-41530 &296.97723 &\-41.88478 & 8.00 &99.60&  7.99& 10& 22.25& 0.50&i& 23.86& 3& V3885-SGR     \\
HST 200803\-48542 &302.01325 &\-48.90438 & 6.18 &99.70&  5.99& 10& 21.00& 0.20&i& 23.74& 3& 2005-489INCA  \\
HST 200811\-48546 &302.04916 &\-48.91062 &15.73 &99.65& 15.20& 20& 22.00& 0.14&i& 23.74& 2& 2005-489INCA  \\
HST 202946+09541 &307.44171 &  9.90304 & 7.00 &99.80&  6.00& 10& 21.25& 0.67&i& 23.76& 2& GL791-2       \\
HST 213233+00161 &323.14022 &  0.26938 & 6.00 &99.50&  6.00& 10& 21.25& 0.39&i& 23.85& 3& LDS749B       \\
HST 214823\-34530 &327.09744 &\-34.88481 & 6.24 &99.99&  5.99& 15& 19.25& 0.70&i& 23.79& 2& IC5135        \\
HST 215031+28488 &327.63226 & 28.81387 & 5.54 &99.84&  5.39& 25& 18.75& 0.40&i& 24.23& 3& BD+28D4211    \\
HST 215112+29002 &327.80405 & 29.00371 &13.24 &99.99&  8.37& 35& 20.00& 0.47&v& 24.84& 3& BD+28D4211    \\
HST 215115+28599 &327.81565 & 28.99922 & 8.37 &99.87&  7.05& 10& 22.00& 0.14&i& 24.17& 4& BD+28D4211    \\
HST 215118+28587 &327.82690 & 28.97907 &10.04 &99.99& 10.02& 25& 22.25& 0.27&b& 25.31& 3& BD+28D4211    \\
HST 215125+29001 &327.85699 & 29.00236 &29.40 &99.99& 25.74& 35& 21.50& 0.43&i& 24.13& 3& BD+28D4211    \\
HST 215128+28581 &327.86881 & 28.96943 &12.00 &99.86& 10.02& 15& 21.75& 0.27&i& 24.13& 2& BD+28D4211    \\
HST 215137+28590 &327.90569 & 28.98339 &14.35 &99.57& 14.00& 15& 22.50& 0.38&i& 24.41& 2& BD+28D4211    \\
HST 225657\-36342 &344.23860 &\-36.57086 &15.85 &99.55& 15.76& 25& 21.50& 0.15&i& 24.56& 1& IC1459-NUC    \\
HST 230425+03051 &346.10657 &  3.08629 & 7.33 &99.60&  7.09& 10& 22.75& 0.50&v& 24.59& 4& PG2302+029    \\ \hline\hline
\end{tabular}
}
\end{table}


}

The MDS cluster richnesses may {\it appear} to be small compared with
the richnesses quoted in the Abell catalogue (see Abell et
al. 1989). However, Abell defined his cluster richnesses within an
aperture of metric radius $1.5h^{-1}$ Mpc. If one scales the observed
surface density of galaxies seen in our MDS clusters/groups to these
larger apertures (using an appropriate cluster profile {\it e.g.}
King), it is clear that some of our clusters would have satisfied
Abell's selection criteria; for example, HST~175525+18182 is equivalent
to an Abell richness class 1 cluster.  We do not provide an ``Abell
richness'' estimate for our clusters in Table 4 because of the
limitations of such an extrapolation. Moreover, we cannot compute the
``Abell richness'' directly as the MDS fields are too small.
 
In column 5, we present the detection probability for each cluster,
while column 6 contains the galaxy richness that corresponds to the
99.5\% cut for that particular aperture ($N_{gal\%}$ in Table 4).
Column 7 is the aperture radius (in arcseconds), while column 8 is the
central magnitude, $m_c$, of the aperture in which the cluster was
detected.  This magnitude, like others used elsewhere in this paper,
is based on the ST system and is an analytical total magnitude derived
from integrating the best fit galaxy model out to 19 half-light radii
(see Ratnatunga et al. 1998).  Column 9 (marked B/T) is the average
for that cluster of the bulge dominated galaxies to the total number
of galaxies within the detection aperture. This average has not been
corrected for contamination by background, or foreground, galaxies and
therefore, must be used care.
 
\begin{figure}[tp]
\centerline{\psfig{file=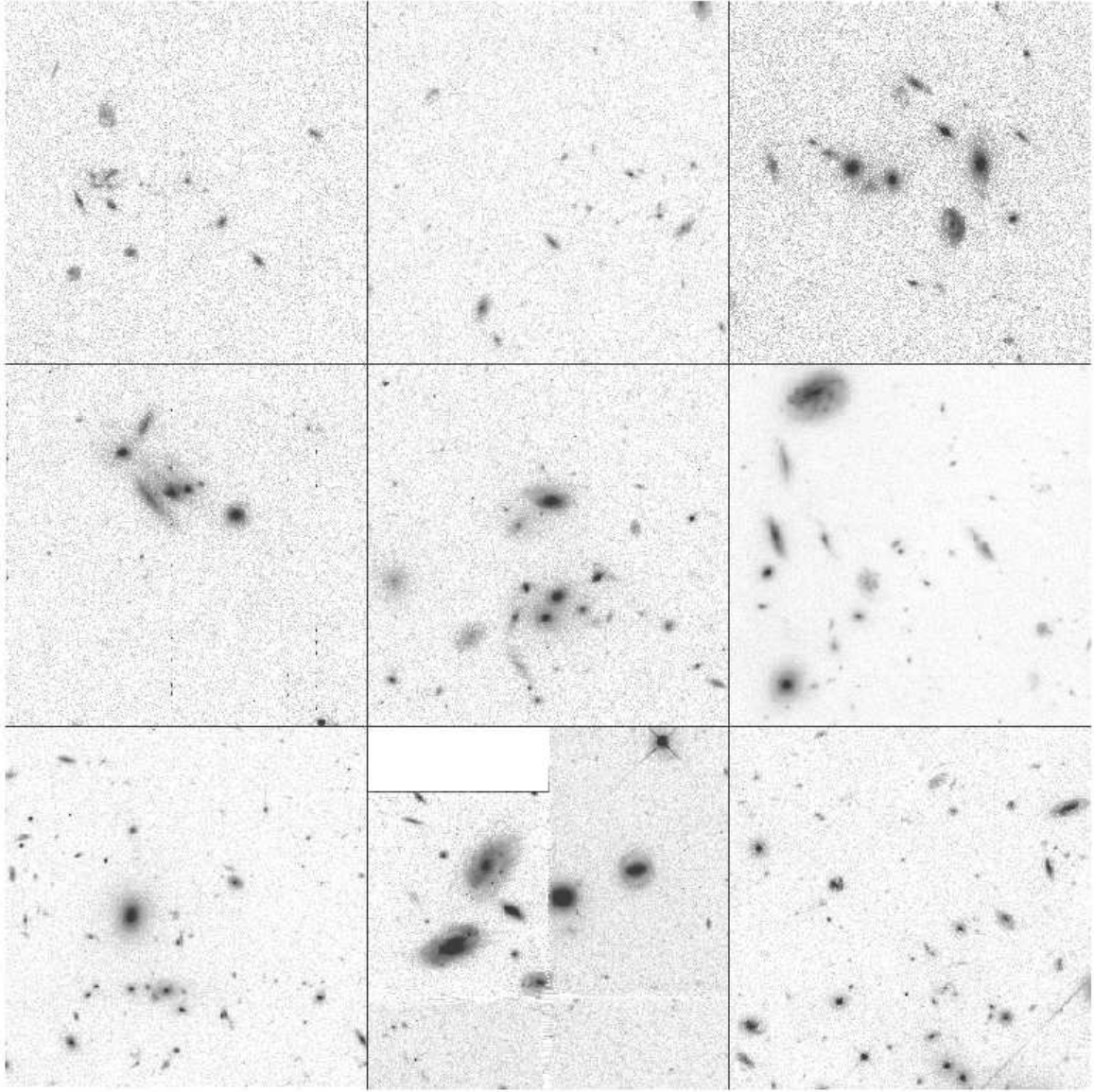,height=8.5in}}
\vspace*{-0.95in}
\caption{
We present here nine examples of clusters and groups found as part of
the HST MDS cluster sample.  The boxes are $30$ arcseconds square and
the passbands are the same as given in Column 10 of Table 4.  On the
top row, from left to right, we present
HST~035528+09435,
HST~115027+28496 and
HST~102722+03257.
On the middle row, from left to right, are
HST~005021-52113,
HST~002013+28366  and
HST~123648+62132.
On the bottom, from left to right, are
HST~141617+52137,
HST~140428+43196 and
HST~133617-00529.
\label{examples}
}
\end{figure}

Column 10 is the WFPC2 filter used and column 11 (marked clim) is the
magnitude completeness limit of the field as defined in Section 2
(Ratnatunga et al. 1998).  Column 12 is a grade between 1 and 4 based
on the visual assessment of the cluster and column 13 is the original
HST target name.  In total, we find 14 clusters classified with grade
1 (``excellent''), 32 as grade 2 (``good''), 29 as grade 3 (``fair'')
and 17 as grade 4 (``poor'').  These subjective assessments are
provided to aid readers in search of a few exceptional clusters for
optical follow--up.  In Fig.~\ref{examples}, we show 9 clusters
selected from Table 4 to highlight the diversity of the catalogue.  In
this figure, we show examples of potentially high redshift systems we
have found, as well as candidates for nearby compact groups of
galaxies.  We also show that clusters can be detected near the edge of
the WFPC2 field--of--view (bottom right) as well as clusters spanning
all 4 WFPC2 chips (bottom center).

In order to estimate photometric redshifts for the galaxy
overdensities presented here, further color information is required as
demonstrated by Connolly et al. (1995).  Estimated redshifts are in
the range from about 0.3 to 1.0, with a mean of 0.7 and a rms of about
0.25.
 
We have cross--correlated the clusters in Table 4 against existing
distant cluster catalogues (Gunn et al. 1986, Couch et al. 1991,
Postman et al. 1996) and found no overlap.  This results from a
combination of our decision to avoid known clusters (see Table 8), and
the fact that the MDS fields are located between 4 and 14 arcmins away
from the primary target (observed with FGS, FOS or FOC).  We also
searched the NASA/IPAC Extragalactic Database (NED) for all possible
matches within the aperture radius of each cluster (given in Table 4).
In total, 5 MDS clusters were coincident with NED sources and these 5
are presented in Table 5 and discussed in more detail below.



\tablenum{5}

\begin{table}[tpb]
\def\-{$-$}
\begin{center}
\caption{The five MDS clusters with a NED coincidence (see text)
\label{nedtable}
}
\begin{tabular}{l|l}\hline\hline
Name & Comments \\ \hline
HST 001831+16208 & Close to RXJ0018.3+1618, RXJ0018.8+1602 and CL0016+16\\ 
                 & all at $z\simeq0.55$.\\
                  & Symmetric double radio source MRC 0015+160,\\
                  & with a 30 arcsecond error circle (95\% confidence),\\
                  & within cluster aperture (see Douglas et al. 1996).\\
                  & Faint X--ray emission associated with cluster \\
                  & and radio source (see Fig. 1 of Hughes et al. 1995).\\
HST 010957\-02276 & Two galaxies within aperture with redshifts $z=0.205\, \&\,z=0.298$. \\
HST 123640+62111 & Flanking the Hubble Deep Field.\\
HST 123649+62132 & Hubble Deep Field. Six redshifts in aperture,\\
                  & three at $z=0.475$. (see text and Fig. 2)\\ 
HST 141727+52267 & Coincident with galaxy CFRS 14.1496 $z=0.899$ \\
                  & (see Lilly et al. 1995).\\
                  & Galaxy CFRS 14.1501 also within aperture radius and is\\
                  & a radio source (Hammer et al. 1995)\\ \hline\hline
\end{tabular}
\end{center}
\end{table}

HST~010957-02276 has a galaxy richness of $5.34$ over a 35 arcsecond
radius aperture and is described as ``poor''. It has two measured
redshifts which are significantly different.
 
HST~123649+62132 is shown in Fig.~\ref{examples} and is at the center
of the HDF.  There are now 6 measured galaxy redshifts within the
aperture radius (from the HDF spectroscopic follow--up, see Cohen et
al. 1996) of which 3 are in agreement; $z=0.475,\, 0.475\, \&\, 0.478$
with all three being ellipticals.  The remaining 3 redshift
measurements are $z=0.199,\, 0.958 \,\&\, 0.749$ with all three
galaxies being spirals or irregulars.  The finding of this overdensity
is thus essentially vindicated, but also illustrates the contaminating
effect of galaxies along the line of sight.
 
\begin{figure}[tp]
\centerline{\psfig{file=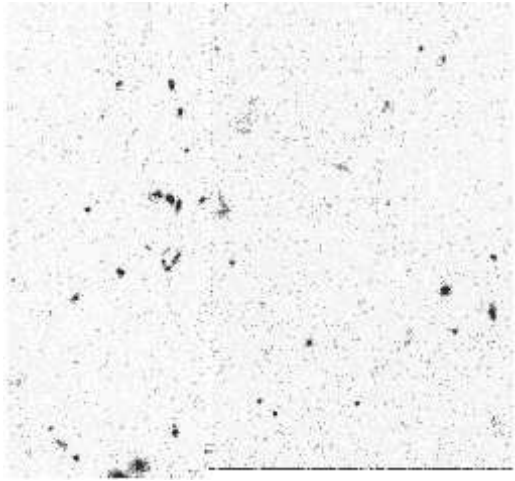,height=2.5in}}
\caption{
This is a greyscale picture of cluster HST~123640+62111 which is
located in the flanking fields of the Hubble Deep Field.  The box is
30 arcseconds square and is in the F814W passband.
\label{MDS2}
}
\end{figure}

HST~123640+62111 was detected in the HST fields that flank the Hubble
Deep Field (see Fig.~\ref{MDS2}).
 
\begin{figure}[tp]
\centerline{\psfig{file=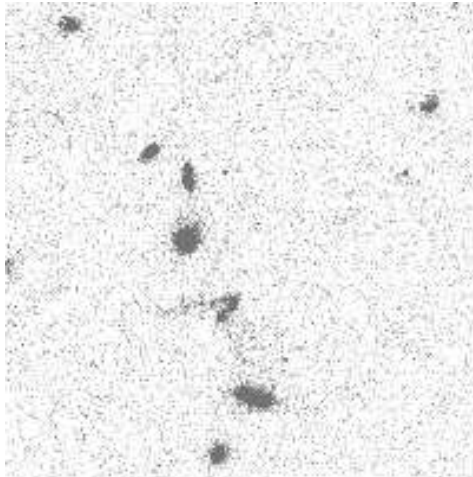,height=2.5in}}
\caption{
This is a greyscale picture of cluster HST~141727+52267 which has one
measured redshift ($0.899$) within it's detection aperture (see text).
The box is 30 arcseconds square and is in the F814W passband.
\label{MDS1}
}
\end{figure}

The central galaxy of HST~141727+52267 is coincident with CFRS 14.1496
at $z=0.899$ from the Canada--France Redshift Survey (CFRS; see Lilly
et al. 1995) and this group is shown in Fig.~\ref{MDS1}. Also within
the aperture is CFRS 14.1501 which is identified as a radio source by
Hammer et al. (1995).
 
HST~001831+16208 is approximately 2 arcmins from RXJ0018.3+1618. This
group, discovered by Hughes, Birkenshaw \& Huchra (1995), is companion
to the famous CL0016+16 cluster at $z=0.546$ (Koo 1981). In addition
to RXJ0018.3+1618, yet another group of galaxies was recently found
close--by, RXJ0018.8+1602 (Connolly et al. 1996), at the same redshift
(Connolly et al. 1997; Hughes \& Birkenshaw 1997) thus raising the
possibility that these clusters/groups are part of a supercluster at
$z=0.55$.  This new MDS cluster potentially adds a further
cluster/group to this high--redshift supercluster.
 
\begin{figure}[tp]
\centerline{\psfig{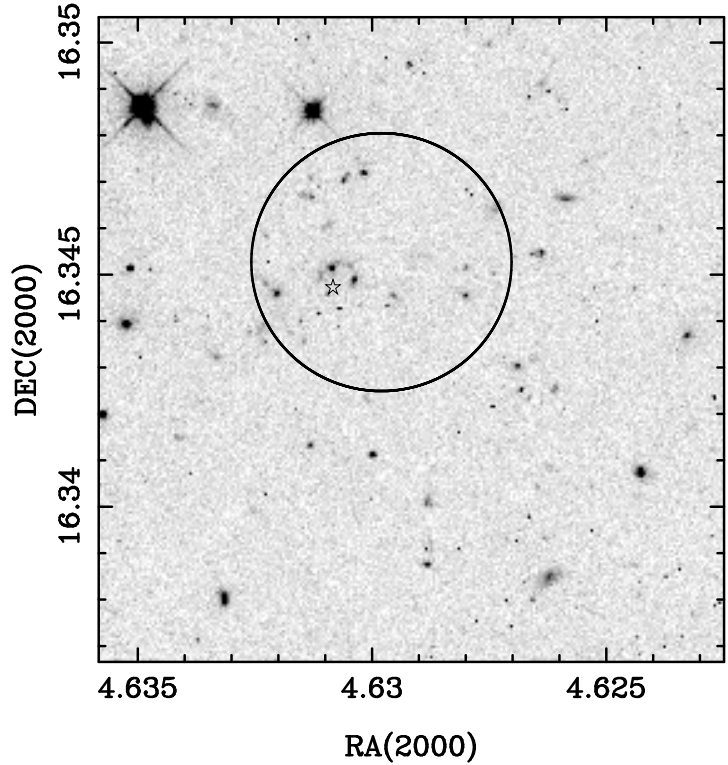}}
\caption{
This is a greyscale picture of HST~001831+16208 in the F814W
filter. The quality of the optical data is poor since this MDS field
received less than 1 hour of exposure time.  The aperture this cluster
was detected in is shown (Table 4) and the position of the known radio
source, MRC 0015+160, is marked with a star.  The image is slightly
rotated.
\label{cl0016}
}
\end{figure}

HST~001831+16208 is a compact group with ${\rm N_{gal}=10.00}$ within a
radius of 10 arcseconds ($\sim75h^{-1}\,{\rm \,kpc}$ diameter at $z=0.6$)
and appears to be coincident with a faint X--ray source as seen in
Figure 1 of Hughes et al. (1995). HST~001831+16208 is very close to the
large cross -- south of the main CL0016+16 cluster -- which marks the
position of a known radio source (MRC 0015+160; see Douglas et
al. 1996).  Figure~\ref{cl0016} shows this new cluster and the
position of the aforementioned radio source.
 
It is worth noting that in addition to the ROSAT PSPC data used by
Hughes et al. (1995), a deep ROSAT HRI pointing towards CL0016+16 also
overlaps HST~001831+16208. Even though the exposure time is longer, the
X-ray source seen in the PSPC data, which is coincident with both
HST~001831+16208 and MRC 0015+160, does not appear to be detected in
the HRI data. Such a non--detection may indicate that the observed
PSPC X--ray emission is slightly extended thus lowering the X--ray
surface brightness of the source below the higher background level in
the HRI data. Follow--up observations are needed to conclusively
determine if this group has associated extended X--ray emission.
 
\section{Simulations}
 
We discuss here the simulations we have performed to assess the likely
contamination of the MDS cluster sample by projection effects {\it
i.e.} how many of the 92 clusters in Table 4 are real and how many are
superpositions of galaxies along the line of sight.  In a future
paper, we will examine the completeness of the survey -- the
percentage of true clusters in the universe detected as a function of
richness, redshift and the selection parameters -- and define the
volume sampled by the MDS cluster sample.
 
To determine the contamination level, we performed two sets of
simulations. The first of these involved taking the MDS galaxy data
and ``shuffling'', at random, the magnitudes of the galaxies.  This
produced fake MDS catalogues with the exact same angular correlations
as the real data, and the same magnitude distributions, but with very
different magnitude correlations. These simulations help to test the
strength of the magnitude correlations for the real clusters/groups.
 
The second set of simulations involved making fake MDS galaxy
catalogues by randomizing the coordinates of the galaxies within each
field.  This ensures that we have the same number and type of galaxies
(bulge and disk dominated) per field and tests the strength of the
observed angular correlations.  We note that these latter simulations
are somewhat naive since they do not fully account for the known
angular galaxy correlations (Collins, Nichol \& Lumsden 1992).
However, with these simulations we are only testing the significance
of galaxy angular correlations within a limited magnitude range (each
aperture is only 2 magnitudes wide).
 
These fake, or simulated, MDS catalogues were searched for clusters
using the same methodology as detailed in the previous section.  The
main difference was that the richness cutoffs, like those given in
Table 3, were not re--computed but the same cutoffs as determined for
the real data were used.



\tablenum{6}

\begin{table}[tb]
\ptlandscape
\begin{center}
\caption{The number of real cluster detections, as a function of $m_c$
and aperture radius, versus the average number of detections in the
simulated MDS cluster catalogs.  [Note: we have used all 126
detections in this table (Section 3.2) as the
multiple detections cannot be removed by hand, unlike the
real MDS catalog which resulted from the reduction of the 126 detections to 92
in Table 4].  The first number within parentheses is the average
number of detections for the shuffled magnitude simulations discussed
in the text. The
second number within parentheses is the average number of
detections for the randomized coordinate simulations.  3000 trials
were run for both sets of simulations.
\label{randomxy}
}
\begin{tabular}{l|rrrrrr} \hline\hline
$m_c$  & \multicolumn{6}{c}{radius arcseconds}\\
     &  10 & 15 & 20 & 25 & 30 & 35 \\ \hline
17.5 &  0 (0.000:0.000)  & 0 (0.000:0.001) & 0 (0.002:0.004) & 1 (1.970:1.098)& 0 (0.264:0.256) & 1 (0.373:0.229)\\
18.5 &  0 (0.000:0.000)  & 0 (0.018:0.028) & 1 (0.072:0.231) & 1 (0.196:0.798)& 1 (0.272:0.946) &  2 (0.367:1.166)\\
19.5 &  2 (0.108:0.132)  & 1 (0.000:0.023) & 3 (0.108:0.278) & 1 (0.216:0.425)& 0 (0.023:0.080) &  4 (0.431:0.869)\\
20.5 &  2 (0.795:1.371)  & 0 (0.132:0.183) & 1 (0.139:0.178) & 2 (0.264:0.289)& 1 (0.134:0.138) &  1 (0.375:0.315)\\
21.5 &  9 (2.154:2.630)  & 1 (0.894:1.136) & 4 (1.013:1.234) & 9 (1.881:2.063)& 5 (1.956:1.918) &  7 (2.356:2.489)\\
22.5 &  16 (5.814:4.443) & 15 (3.195:2.279) & 5 (3.885:3.591) & 4 (3.090:2.693)& 0 (2.313:2.186) & 15 (7.311:7.333)\\
23.5 &  3 (3.469:1.749)  & 2 (1.327:0.706) & 0 (1.283:0.738) & 1 (0.542:0.268)& 1 (0.349:0.363) &  0 (1.291:0.318)\\
24.5 &  1 (0.121:0.514)  & 0 (0.162:0.406) & 2 (0.215:0.630) & 0 (0.000:0.000)& 0 (0.000:0.000) & 0 (0.000:0.000)\\\hline\hline
\end{tabular}
\end{center}
\end{table}

In Table 6, we present the results of these simulations based on
searching the fake MDS catalogues (3000 for both sets of simulations).
This table shows the real number of cluster detections, as a function
of their detection aperture radius and $m_c$, together with the
average number of detected clusters in the fake catalogues (we have
coarsely re--binned the values of $m_c$ into bins of width 1
magnitude).  This helps to demonstrate the expected rate of
contamination as a function of these two detection parameters and will
help the reader to determine the significance of any particular
cluster detection compared to others. For example, in Table 6 the two
clusters detected at $19<m_c<20$ and ${\rm radius}=10''$ (0.108:0.132)
are much more likely clusters than the two detected at $23<m_c<24$ and
${\rm radius}=15''$ (1.327:0.706) because of the relative increase in
the average number of false detections in both sets of simulations
(given in brackets).
 
A close examination of Table 6 shows that the number of fake clusters
detected as a function of $m_c$ and aperture radius is correlated with
the actual number of detections in that bin. At bright magnitudes this
is due to the fact that a real cluster in the data can add a
significant number of galaxies to that magnitude bin and therefore, it
will be detected again and again in the fake datasets. In such a case,
the average number of fake detections in these bright magnitude bins
is overestimated compared with a truly random process. At fainter
magnitudes, the volume of the survey is increasing along with the
number of galaxies in the MDS. Therefore, one would expect to find
more clusters in both the real dataset and in the fake catalogues.
 
In addition to simulating our false detection rate, we also carried
out simulations to test our cluster finding algorithm and our original
hypothesis that centering on bulge dominated galaxies increases our
success rate in detecting real clusters/groups.  For these extra
simulations, we again computed the ratio of real galaxy overdensities
detected in the MDS data versus the average number of fake clusters
detected in the simulated catalogues (both shuffled magnitudes and
randomized coordinates).  The crucial difference for these extra
simulations was that we centered on disk dominated galaxies, as
opposed to bulge dominated galaxies above, and then repeated the
exercise by centering on all galaxies regardless of their
morphological classification.



\tablenum{7}

\begin{table}[bt]
\begin{center}
\caption{A summary of all simulations carried out to understand 
our cluster--finding algorithm. The left--hand column presents
the type of simulation carried out --  magnitude--shuffling
or the randomizing of the coordinates -- as well as the HST
passband used. 
\label{sumsim}
}
\begin{tabular}{lccc}\hline\hline
Type of     & \multicolumn{3}{c}{Morphology}\\
simulation  &  All Galaxies & Bulge & Disk \\ \hline
F450W \& magnitudes & 54\% & 34\% & 60\% \\
F450W \& coordinates& 71\% & 31\% & 83\% \\
F606W \& magnitudes & 51\% & 44\% & 50\% \\
F606W \& coordinates & 48\% & 40\% & 48\% \\
F814W \& magnitudes & 53\% & 40\% & 55\% \\
F814W \& coordinates & 45\% & 38\% & 46\% \\ \hline\hline
\end{tabular}
\end{center}
\end{table}

In Table 7, we present a summary of all these simulations. The table
shows the type of simulation carried out, magnitude shuffling or
coordinate randomization, along with the bandpass used and the
morphology of the centered galaxy {\it i.e.} bulge dominated, disk
dominated or all galaxies regardless of the morphological
classification. The percentages given in this table are the ratios of
the average number of detections in these simulations compared to the
number of detections in the real overdensity database.  Table 7
demonstrates that by centering on bulge dominated galaxies, the
overall expected false--detection rate should be between 30\% to 40\%;
in other words, {\it at least} 60\% of the MDS clusters are likely
real overdensities. This is smaller than the false--detection rates
estimated by van Haarlem et al. (1997) for the Abell catalogue, but
comparable to the estimate of Lucey (1983).
 
Table 7 also shows that the false detection rate increases
substantially for the other methodologies {\it i.e.} centering on all
galaxies as opposed to just bulge dominated galaxies. The worst case
is centering on disk dominated galaxies where the percentage of false
detections, compared to real detections, is $>50\%$. It is not
surprising that the numbers given in Table 7 for all galaxies closely
follows those given for disk dominated galaxies as 80\% of all MDS
galaxy morphologies are disk dominated.
 
\section{Selection Function}
 
The methodology outlined in this paper allows us to determine the
selection function of the MDS cluster sample.  This is an important
quantity as it will enable us to estimate the sampled volume.  We
present here an empiricial discussion of the selection function.
 
One of the main caveats of the MDS cluster sample, as presented in
Table 4, is that it has been selected using morphological
criteria. This may be unsuitable at high redshift if the morphologies
of cluster galaxies are shown to evolve.  Moreover, our algorithm
implicitly assumes that the angular galaxy correlation functions for
bulge dominated and disk dominated galaxies are the same, which may
not be true.  Our simulations have shown that our morphology--based
methodology has helped to increase the ratio of real clusters detected
compared with spurious systems (see Table 7).  However, to alleviate
some of these concerns, we have made available via the MDS
World--Wide--Web homepage (http://astro.phys.cmu.edu/mds/), a complete
list of detected overdensities regardless of the morphology of the
central galaxy. The reader is warned that there is a higher level of
contamination in this catalogue (see Table 7). The original galaxy
catalogues used in the search (in all three passbands) are made
available, together with all 126 clusters detected after the first
pass through the candidates following removal of duplicates (see
Section 3.2) {\it i.e.} we provide the 34 clusters ($126-92$) that
were altered by hand as we believed they were also duplicates.  Also
presented are the galaxy richness cutoffs for the F450W and F606W
passbands (like those given in Table 3), a copy of the computer
software (fortran) used and a list of the MDS fields searched.  These
data are provided to facilitate checks of our clusters and
methodology.
 
Another potential bias with the HST MDS cluster survey is the small
field--of--view of the WFPC2 instrument.  Clusters of galaxies are
large objects on the sky and typically the core radius of clusters is
larger than the WFPC2 field size; for example, at $z=1.25$, near the
minimum in the angular diameter--redshift relation for currently
popular cosmological models, a typical cluster core radius
($250h^{-1}$kpc) subtends $\sim1$ arcmin on the sky which is larger
than any aperture we use in our analysis. Therefore, the reader is
warned that the HST MDS cluster survey is certainly biased towards
compact systems.
 
To investigate this potential bias, we processed HST WFPC2 archival
data on known distant clusters using the same algorithm as described
above.  We did not redetermine the thresholds, but used the same ones
as discussed above and presented in Table 3. We also sorted the
clusters and removed overlapping systems as discussed in Section 3.2.
The cluster fields used in this analysis are shown in Table 8 where we
present the MDS field identifier, the name of the target cluster, the
name of the Principal Investigator and the redshift of the cluster
(taken from the NED database). We also include the number of cluster
detections in each of these fields using our MDS cluster--finding
algorithm.  We note that this number of cluster detections still
includes some multiple detections because we have not merged the
cluster lists from different passbands. Therefore, it is an upper
limit and should be used for illustrative purposes only.

%


\tablenum{8}


{\small
\begin{table}[tp]
\def\-{$-$}
\begin{center}
\caption{WFPC2 archival data towards known clusters of galaxies.
We present the (internal) MDS field identifier, the primary target name,
the name of the Principal Investigator, the number of HST pointings towards
the cluster, a redshift (if known) from the 
NED database, any alternative name for the cluster, and 
the number of detections for this cluster using the HST MDS cluster
survey algorithm.
\label{clusters}
}
\begin{tabular}{c|lrlrlr}\hline\hline
MDS Field & Cluster Name       & No. of    & Name of PI & Redshift  & Comments & No. \\
Identifier& (in HST Arcive)    & Pointings &            &  (if known) & (alt. names) & Detections \\ \hline
u2841--u2842 & GAL-CLUS1322+3027 &2& Westphal   & 0.751 &GHO 1322+3027 &15\\
u2845 & GAL-CLUS1603+4313 &1& Westphal   & 0.895 &GHO 1602+4312 & 1 \\
u29g1--u29g2 & GAL-CLUS-002635+170944 &2& Turner  & 0.390 & ZwCl 0024.0+1652 &40 \\
u2c41 & GAL-CLUS-001558+1609   &1& Dressler   & 0.541 & Cl0016+16 & 25\\
u2c42 & GAL-CLUS-005431\-2756 &1&   Dressler   & 0.56 & J1888.16CL & 13 \\
u2c44 & GAL-CLUS-041234\-6558 &1& Dressler & 0.51 & F1557.19TC & 25\\
u2c47 & GAL-CLUS-093942+47 &1& Dressler & 0.402 & A851 & 16 \\
u2c48 & A370 &1& Dressler & 0.373 & & 14 \\
u2fq1 & A2390 &1&  Fort & 0.231 & RX J21535+1741 & 31 \\
u2fq2 & CL2244-02 &1&  Fort & 0.330 & EXSS 2244.6\-0220 & 16 \\
u2gk1--u2gk2 & A665 &2& Franx & 0.181 & & 43\\
u2uj1--u2ujb & GAL-CLUS-135951+623105 &10& Franx & 0.328 & MS 1358.4+6245 & 84 \\
u2ul1 & GAL-CLUS-030533+171005 &1& Illingworth & 0.424 &MS 0302.7+1658 & 5\\
u2ul2 & GAL-CLUS-030518+172838 &1& Illingworth &0.425 & MS 0302.5+1717 & 9\\
u2ul4 & GAL-CLUS-214012\-233927 &1& Illingworth &0.313 & MS 2137.3\-2353 & 13 \\
u2um1 & GAL-CLUS-073924+702311 &1& Fevre &  & & 2 \\
u2vk2 & GAL-CLUS-023143+004844  &1&  Postman &  & &2 \\
u2w91--u2w98 & A1689  &4& Kaiser &0.181 & &43\\
u3062 & MS1054\-03 &1& Donahue & 0.823 & &26 \\
u3063 & MS1137+66 &1& Donahue & 0.782 & &34 \\
u30h1 & 3C215 &1& Ellingson & 0.411 & QSO field & 6\\
u30h2 & 5C2.10 &1& Ellingson & 0.478 & QSO field & 14 \\
u30h3 & 3C281 &1& Ellingson &  & QSO field & 5 \\
u34e1 & GAL-CLUS1322+3115 &1& Westphal &0.755 &GHO 1322+3114 & 13 \\
u34m2 & FIELD-0336\-3645 &1& Grillmair & & Fornax $z=0.0046$ & 1\\
u34m3 & FIELD-0338\-3523 &1& Grillmair & & Fornax $z=0.0046$ & 4\\ \hline\hline
\end{tabular}
\end{center}
\end{table}
}

The fields in Table 8 were excluded from the main MDS cluster survey
analysis because they would have biased the serendipitous nature of
our survey as well as dominating our statistical distributions because
of the sheer number of cluster detections; these cluster fields are
significantly overdense compared to the normal MDS fields and as a
result, we find hundreds of candidate clusters in these fields. In
Table 9, we give the average number of clusters/groups detected, as a
function of passband, in both the targetted cluster fields (Table 8)
and the main MDS fields. On average, we detect $\sim25$ times more
clusters/groups in known cluster fields than in general MDS fields;
this is the result of the fact that every elliptical in these known
clusters produces a candidate MDS cluster.  This exercise demonstrates
that our algorithm is very sensitive to a wide range of clusters, both
in redshift, $\sim0.2< z {<\atop{\sim}} 0.9$, and richness {\it i.e.} from A665, which is
one of the richest clusters in the Abell catalog, to the proposed
group around 3C215 (Ellingson \& Yee 1994).  Therefore, we may not be
severely biased against clusters which are much larger than the WFPC2
instrument {\it i.e.} low redshift systems.  If one of the MDS fields
had accidentally fallen upon such a low-redshift cluster, Tables 8 \&
9 demonstrate that we would have detected it.
 
%


\tablenum{9}


\begin{table}[tb]
\begin{center}
\caption{The number of clusters detected in WFPC2 fields towards 
known clusters of galaxies (Table 8).
We give the passband used in the analysis, the number of
cluster fields exposed to that passband, the total number of clusters
detected in those fields and the average number of clusters
detected per field (labelled Av. detection rate 1). 
For comparison, we present the average
number of clusters detected per field in the main MDS cluster
survey (labelled Av. detection rate 2).
\label{clusterfields}
}
\begin{tabular}{lcccc}\hline\hline
Filter & No. Cluster & No. Clusters & Av. Detection  & Av. Detection\\
       & Fields      & Detected     & Rate 1         & Rate 2        \\ \hline
F450W (b) & 4  & 26  & 6.5 & 0.4 \\
F606W (v) & 36 & 280 & 7.8 & 0.3 \\
F814W (i) & 43 & 388 & 9.0 & 0.4 \\ \hline \hline
\end{tabular}
\end{center}
\end{table}


We note here that the data presented in Tables 8 \& 9 is
scientifically interesting beyond simple tests of our algorithm. This
work allows us to put known, well studied clusters into the same
statistical framework as the rest of the MDC Cluster Survey. For
example, it allows us to assess the statistical significance of any
clusters found in these fields compared to the cluster population as a
whole.
 
\begin{figure}[tp]
\centerline{\psfig{file=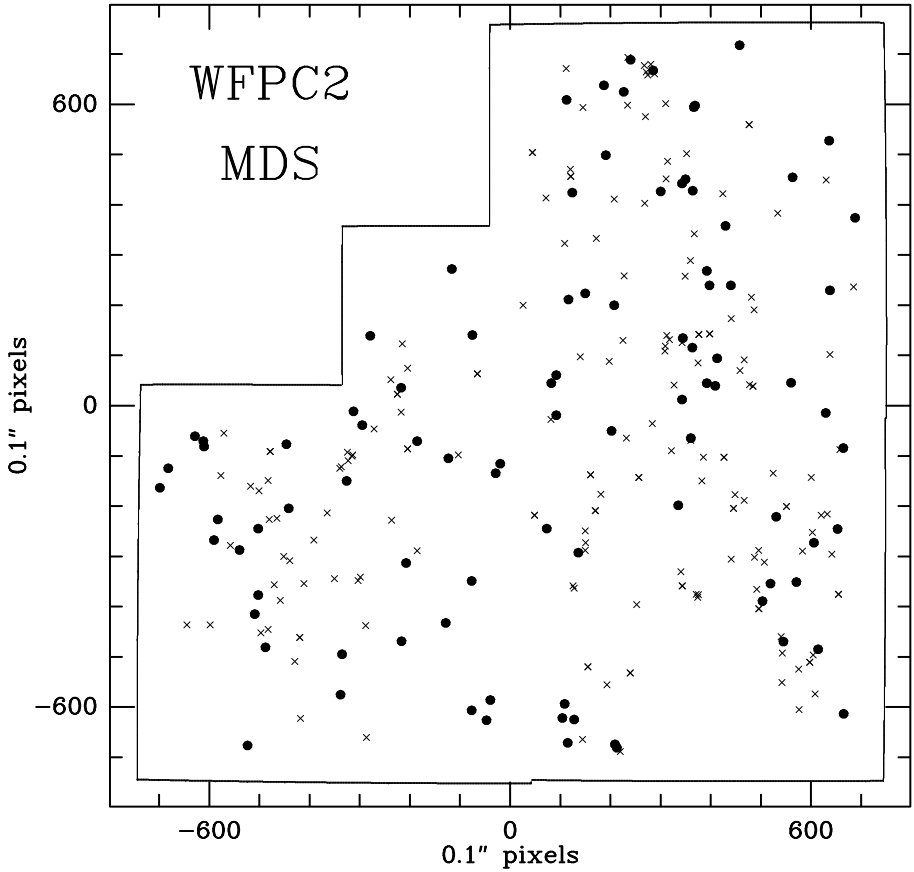,height=4.5in,angle=270}} 
\caption{
The pixel coordinate of cluster detections within the WPFC2
field--of--view. Shown are the 3 main CCD chips as well as the
Planetary Camera in the top-left quadrant.  The $\bullet$ symbols are
all sorted cluster detections in the F814W data, {\it i.e.} removal of
overlapping systems, while the $\times$ symbols are the unsorted F814W
detections.  The thin soild line outlines the nominal boundary of the
WPFC2 instrument (a total area of $4.75{\rm arcmins^2}$. As expected,
we are insenstive to cluster detections close to the edges of the
field--of--view (we demand that 75\% of the detection aperture is
within the MDS field). We have computed the effective cluster search
area of a WPFC2 field using this figure and it is $4.27{\rm
arcmins^2}$.
\label{plotarea}
}
\end{figure}

Finally, in Figure~\ref{plotarea} we show the distribution of cluster
detections plotted as a function of their WFPC2 instrumental
coordinates.  This exercise demonstrates that our algorithm detects
cluster candidates evenly throughout the WFPC2 field; it is not biased
towards preferentially finding clusters near the field center. The
algorithm finds clusters in the corners of the field as well as across
CCD boundaries.  This exercise also allowed us to determine the
effective area of the WPFC2 instrument to our cluster search.
Accounting for the decrease in sensitivity towards the edges of the
field while including the Planetary Camera, the effective search area
was computed to be 4.27 ${\rm arcmins^2}$.

\section{Discussion and Conclusions}
 
In this paper, we present the results of a search for galaxy
overdensities within the HST Medium Deep Survey. This search was fully
automated and has an objective selection function. The 92 clusters
found are presented in Table 4 and are the richest, most significant
overdensities within the MDS survey.
 
A close examination of Table 4 demonstrates that we have potentially
found some very high redshift clusters/groups, with about 25\% of the
sample having estimated redshifts $z {>\atop{\sim}} 0.9$. If we
restrict ourselves to considering only candidates rated as
``excellent'' or ``good'', then we still have about 10\% of our
systems above this redshift.  This significantly increases the number
of potential cluster candidates above this particular redshift, since
the PDCS has only 9 cluster candidates at $z_{est}>0.9$.  These
combined optical catalogues may provide excellent targets for studies
of cluster and galaxy evolution.
 
We note here that the PDCS covers an area of $5 {\rm deg^2}$, a factor
of 10 greater than the MDS HST cluster sample, yet we find a
comparable number of $z>0.9$ clusters. This apparent discrepancy in
the surface density of such systems should not be over interpreted, as
it is simply a reflection of the different photometric limits of the
two surveys (PDCS is complete to only $I_4=22$). Also, it is a
reflection of the different cluster--finding algorithms: Postman et
al. did not attempt to search for clusters below an Abell richness
class 1 at such high redshift, so the PDCS is only complete in the
redshift range $0.2<z<0.6$.
 
\begin{figure}[tp]
\centerline{\psfig{file=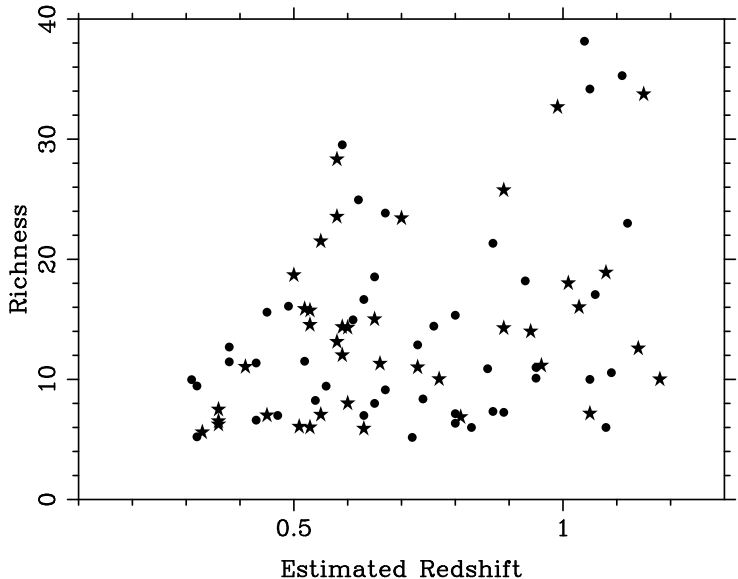,height=3.in,angle=270}} 
\vspace*{0.5in}
\caption{
The richness of the MDS clusters versus their estimated redshifts. The
$\star$ symbols are clusters classified as ``excellent'' or ``good''
(see Table 4), while the $\bullet$ symbols are the remaining clusters.
\label{rich}
}
\end{figure}

In Figure~\ref{rich}, we show the richness of our systems versus
redshifts roughly estimated using $(V - I)$ color, following the
prescription of Im (95). This figure indicates that at lower
redshifts, we are only detecting poor systems {\it i.e.} groups. This
is to be expected since the volume sampled at these lower redshifts is
small and therefore, we are insensitive to rich clusters which have a
low space density. At higher redshift, however, we do appear to be
detecting richer systems as our volume increases. One caveat to this
statement is that at higher redshift the amount of line--of--sight
contamination increases as well.

We have performed extensive Monte Carlo simulations to help understand
our cluster--finding algorithm and estimate our false--detection rate.
These simulation have shown that $>60\%$ of the MDS cluster sample are
real spatial overdensities and not due to chance projections on the
sky.  This percentage is in good agreement with our visual impression
of these systems where we have classified 50\% of the MDS clusters as
either ``excellent'' or ``good'' (see Table 4).  Our simulations also
show that this percentage decreases if one widens the search for
clusters to disk dominated galaxies or all galaxies.  This
demonstrates that our methodology does help to minimize the rate of
false detections and increases our sensitivity to real
groups/clusters.
 
We cross--correlated the MDS cluster sample against known catalogues
of extragalactic objects including previous catalogues of distant
clusters (Gunn et al. 1986; Couch et al. 1991; Postman et al. 1996).
Only five MDS clusters had a coincidence with a known source, none of
which were known clusters.  This demonstrates that most of our MDS
clusters are new, previously unknown systems. Both HST~010957-02276 and
HST~123648+62132 have multiple galaxy redshift measurements which
indicate that the former is probably spurious while the latter cluster
is a bound group (in the HDF field).  HST~141727+52267 has one galaxy
redshift measurement of $z=0.899$, while HST~001830+16208 is close on
the sky to the massive cluster CL0016+16 thus supporting the
hypothesis that these clusters and groups are part of a supercluster
at $z=0.55$ (Connolly et al. 1996).
 
The MDS cluster sample will be useful to address key issues in cluster
research.  First, in Table 4, we present the ratio of the bulge
dominated galaxies in our systems compared to the total number of
galaxies.  This ratio indicates the overall morphological constitution
of these overdensities and, in future, can be used to probe the
evolution of clusters/groups as a function of epoch {\it i.e.} to help
address the Butcher \& Oemler (1984) effect and the
density--morphology relation (Dressler 1980).  However, such analyses
will require redshift measurements before they can be conclusive.
 
Finally, the volume of the MDS cluster sample can be estimated via
extensive Monte Carlo simulations, thus allowing the number density of
our systems to be calculated.  This may possibly provide a constraint
on structure formation models and on the value of $\Omega$ (see
Blanchard \& Bartlett 1997, Reichart et al. 1998).
 
\section{Acknowledgements}
 
The data presented here are from the HST Medium Deep Survey project,
and we acknowledge the other members of the MDS team (see author list
in Griffiths et al. 1994).  This work is based on observations taken
with the NASA/ESA {\it Hubble Space Telescope} obtained at the Space
Telescope Science Institute, which is operated by the Association of
Universities for Research in Astronomy, Inc. The HST Medium Deep
Survey has been funded by STScI Grants GO 2684.0X.87A and GO
3917.0X.91A, and we also acknowledge support from STScI Grant GO 7536
for archival analysis of clusters. We thank Kathy Romer, Jeff Peterson
and Andy Connolly for helpful discussions about this work and for
constructive comments on earlier drafts of this paper.  We are
grateful to Jack Hughes for discussions concerning the ROSAT HRI data
towards HST~001831+16208.  We thank the referee, Francisco Castander,
for his constructive comments which resulted in many improvements to
the paper.  EJO and KUR were partially supported by JPL contract
959501 for WFPC2 GTO team science and RCN was partially supported on
NASA grant NAG5-6548.  This research has made use of the NASA/IPAC
Extragalactic Database (NED) which is operated by the Jet Propulsion
Laboratory, California Institute of Technology, under contract with
the National Aeronautics and Space Administration.


\begin{thebibliography}{}
 
\bibitem[Abell 1958]{abell56}
Abell, G. O., 1958, \apjs, 3, 211.
 
\bibitem[Abell 1958]{ab56}
Abell, G. O., Corwin, H. G. \& Olowin, R. P., \apjs, 70, 1.
 
 
\bibitem[Blanchard et al. 1986]{ball86}
Blanchard, A., \& Bartlett, J. G.,
1997, A\&A, 332L, 49.
 
\bibitem[Bardeen et al. 1986]{but86}
Butcher, H. R., \& Oemler, A., 1978, \apj, 219, 18.\
 
\bibitem[Bardeen et al. 1986]{cohen86}
Cohen, J. G., Cowie, L. Hogg, D. W., Songaila, A., Blanford, R., Hu, E. M.,
\& Shopbell, P., 1996, \apj, 471, L5.
 
\bibitem[Bardeen et al. 1986]{collins86}
Collins, C. A., Nichol, R.C, \& Lumsden, S. L., 1992, \mnras, 254, 295.
 
\bibitem[connolly et al. 1986]{con96}
Connolly, A. J., Csabai, I., Szalay, A. S., Koo, D. C., Kron, R. G.
\& Munn, J. A., 1995, AJ, 110, 2655
 
\bibitem[connolly et al. 1986]{con86}
Connolly, A. J., Szalay, A. S., Koo, D. C., Romer, A. K.,
Holden, B. P., Nichol, R. C., \& Miyaji, T., 1996, \apj,
473, L69
 
\bibitem[connolly et al. 1986]{con87} 
Connolly, A. J., Szalay, A. S., Romer, A. K., Nichol, R. C., Holden,
B. P., Koo, D. C., \& Miyaji, T., 1997, in Proceedings of the 18th
Texas Symposium in Relativistic Astrophysics, in
press. Astro--ph/9702025.
 
\bibitem[Couch et al. 1991]{couch86}
Couch, W. J., Ellis, R. E., Sharples, R. M., \& MacLaren, I., 1991,
\mnras, 249, 606.
 
\bibitem[Gioia \& Luppinno]{dick86}
Dickinson, M., 1997, Proceedings of ESO/VLT meeting ``Galaxy Scaling
Relations: Origin, Evolution and Applications''. ed. L. da Costa.
Astro-ph/9703035
 
\bibitem[Bardeen et al. 1986]{doug96}
Douglas, J. N., Bash, F. N., Bozyan, A. F., Torrence, G. W., \& Wolfe
C., 1996, \aj, 111, 1945.
 
\bibitem[Bardeen et al. 1986]{dress86}
Dressler, A., 1980, \apj, 236, 351.
 
\bibitem[Bardeen et al. 1986]{dress86}
Ellingson, E. \& Yee, H. K. C. 1994, ApJS, 92, 33.
 
\bibitem[Couch et al. 1991]{frenk86}
Frenk, C. S., White, S. D. M., Efstathiou, G., \& Davis, M.,
1990, \apj, 351, 10.
 
\bibitem[Griffiths et al. 1994]{g94}
Griffiths, R. E., Ratnatunga, K. U., Neuschaefer, L .W.,
Casertano, S., Im, M., Wyckoff, E. W., Ellis, R. E., Gilmore, G. F.,
Elson, R. A. W., Glazebrook, K., Schade, D. J., Windhorst,
R. A., Schmidtke, P. C., Gordon, J. M., Pascarelle, S. M.,
Illingworth, G. D., Koo, D. C., Bershady, M. A., Forbes, D. A.,
Phillips, A. C., Green, R. F., Sarajedini, V., Huchra, J. P.
\& Tyson, A. J., 1994, \apj, 437, 67.
 
\bibitem[Gunn et al. 1991]{gunn86}
Gunn, J. E., Hoessel, J. G., \& Oke, J. B., 1986, \apj, 306, 30.
 
\bibitem[huchra et al. 1990]{hammer92}
Hammer, F., Crampton, D., Lilly, S. J., Le Fevre, O., \& Kenet, T., 1995, \mnras, 276, 1085.
 
\bibitem[hickson 1997]{hic97}
Hickson, P., 1997, \araa, 35, 357.
 
\bibitem[huchra et al. 1990]{hu9}
Huchra, J. P., Henry, J. P., Postman, M., \& Geller, M. J., 1990,
\apj, 365, 66.
 
\bibitem[huchra et al. 1990]{huc92}
Hughes, J. P., Birkenshaw, M., \& Huchra, J. P.
1995, \apj, 448, L93
 
\bibitem[huchra et al. 1990]{hughes92}
Hughes, J. P., \& Birkenshaw, M.,
1997, \apj, 497, 645.
 
\bibitem[huchra et al. 1990]{hugh92}
Im, M., 1995, Ph.D thesis, Johns Hopkins University
 
\bibitem[huchra et al. 1990]{hu92}
Im, M., Griffiths, R. E., Ratnatunga, K . U., \& Sarajedini, V.,
1996, \apj, 461, L79
 
\bibitem[huchra et al. 1990]{kas92}
Kawasaki, M., Shimasaki, K., Doi, M., \& Okamura, S., 1997,
A\&AS, 130, 567.
 
\bibitem[kepner et al. 1998]{kepner}
Kepner, J., Fan, X., Bahcall, N., Gunn, J. E. \&  Lupton, R., 1998,
ApJ, astro-ph/9803125
 
\bibitem[kodama et al. 1990]{k00ii}
Kodama, T., Bell, E. F. \& Bower, R. G., 1998, MNRAS, astro-ph/9806120
 
\bibitem[huchra et al. 1990]{k0081}
Koo, D. C., 1981, \apj, 251, L75
 
\bibitem[huchra et al. 1990]{lid92}
Lidman, C. E., \& Peterson, B. A., 1996, \aj,
112, 2454
 
\bibitem[huchra et al. 1990]{lily92}
Lilly, S. J., Hammer, F., Le Fevre, O., \& Crampton, D., 1995, \aj, 455, 75.
 
\bibitem[huchra et al. 1990]{lucey92}
Lucey, J. R., 1983, \mnras, 204, 33.
 
\bibitem[Nichol et al. 1992]{lum}
Lumsden, S. L., Nichol, R.C, Collins, C. A.,\& Guzzo, L. 1992,
\mnras, 258, 1.
 
\bibitem[Postman et al. 1992]{post92}
Postman, M., Huchra, J. P. \& Geller, M. J., 1992, \apj, 384, 404.
 
\bibitem[Postman et al. 1996]{post96}
Postman, M., Lubin, L. M., Gunn, J. E., Oke, J. B.,
Hoessel, J. M., Schneider, D. P., \& Christensen, J. A.,
1996, \aj, 111, 615.
 
\bibitem[Press \& Schechter 1974]{PS74} Press, W. H, \& Schechter, P., 1974,
\apj, 187, 425.
 
\bibitem[Ratnatunga et al. 1994]{rat94}
Ratnatunga, K. U., Griffiths, R. E., Casertano, S., Neuschaefer L. W.
\& Wyckoff, E. W., 1994, \aj, 108, 2362.
 
\bibitem[Ratnatunga et al. 1998]{rat98}
Ratnatunga, K. U., Griffiths, R. E., Ostrander, E. J., 1998, in preparation
(see http://astro.phys.cmu.edu/mds/mle/)
 
\bibitem[Ratnatunga et al. ]{rat8}
D. E. Reichart, D.E., Nichol, R.C., Castander, F.C., Burke, D., Romer, A.K.,
Holden, B.P., Collins, C.A. \& Ulmer, M.P., ApJ, astro-ph/9802153
 
\bibitem[Richstone et al. 1990]{rich90}
Richstone, D., Loeb, A. \& Turner, E. L., 1992, \apj, 393, 477.
 
\bibitem[Roche et al. 1992]{ric92}
Roche, N., Ratnatunga, K. U., Griffiths, R. E. \& Im, M., 1998,
\mnras, 288, 200
 
\bibitem[Richstone et al. 1990]{seee90}
Slezak, E., Bijaoui, A \& Mars, G., A\&A, 227, 301
 
\bibitem[Richsone et al. 190]{see90}
Williams, R. E. et al. 1996, AJ, 112, 1335
 
\bibitem[Richstone et al. 1990]{van90}
van Haarlem, M. P., Frenk, C. S., \& White, S. D. M.,
1997, \mnras, 287, 817.
 
\bibitem[zwicky et al. 1968]{zzz90}
Zwicky, F., Herzog, E., Wild, P., Karpowicz, M.
\& Kowal, C. T. 1968, Catalogue of Galaxies and Clusters
of Galaxies (California Institute of Technology, Pasadena)
 
\end{thebibliography}
\end{document}